\documentclass[aps,twocolumn,prl,showpacs]{revtex4}
\usepackage{graphicx}
\usepackage{amsmath,amssymb,amsthm,bbm,latexsym}
\usepackage{amsthm}
\usepackage{amsbsy}
\usepackage{epsfig}
\usepackage{graphicx}
\usepackage{float}
\usepackage{dcolumn}
\usepackage{amsmath}

\begin{document}

\title{Bounds on Negativity of Superpositions}
\author{Yong-Cheng Ou and Heng Fan }
\affiliation{Institute of Physics, Chinese Academy of Sciences, Beijing 100080, People's
Republic of China}

\begin{abstract}
The entanglement quantified by negativity of pure bipartite superposed
states is studied. We show that if the entanglement is quantified by the
concurrence two pure states of high fidelity to one another still have
nearly the same entanglement. Furthermore this conclusion can be guaranteed
by our obtained inequality, and the concurrence is shown to be a continuous
function even in infinite dimensions. The bounds on the negativity of
superposed states in terms of those of the states being superposed are
obtained. These bounds can find useful applications in estimating the amount
of the entanglement of a given pure state.
\end{abstract}

\maketitle

Quantum entanglement plays an important role both in many aspects of quantum
information theory\cite{nielsen} and in describing quantum phase transition
in quantum many-body systems\cite{os, oo}. As such characterization
quantification of quantum entanglement is a fundamental issue. Consequently
the legitimate measures of entanglement are desirable as a first step. The
existing well-known bipartite measure of entanglement with an elegant
formula is the concurrence derived analytically by Wootters\cite{wootters}
and the entanglement of formation\cite{bennett, hill} is a monotonically
increasing function of the concurrence. In general for a multipartite or
higher-dimensional system it is a formidable task of quantifying its
entanglement since it needs complicate convex-roof extension. In the last 10
years some important properties of quantum entanglement were found, one of
which is the monogamy property described by Coffman-Kundu-Wootters
inequality in terms of concurrence \cite{coffman}. In our previous work we
have shown that the monogamy inequality can not generalize to
higher-dimensional systems\cite{ou1} and established a monogamy inequality
in terms of negativity giving a different residual entanglement\cite{ou2}.

On the other hand, quantum entanglement is a direct consequence of the
superposition principle. It is an interesting physical phenomenon that the
superposition of two separable states may give birth to an entangled state,
on the contrary, the superposition of two entangled states may give birth to
a separable state. The relation between the entanglement of the state and
the entanglement of the individual terms that by superposition yield the
state has been studied, where the entanglement is quantified by the von
Neumann entropy\cite{linden} and the concurrence\cite{yu}. Recently it was
generalized to the superposition of more than two components\cite{yang}. If
the entanglement is quantified by negativity, it would be interesting to
establish the analogous relation and obtain the bound of entanglement for
the superposition state. In this paper, we first show that, by contrast to
the von Neumann entropy, the concurrence is a continuous function even in
infinite dimensions. We deduce an inequality to guarantee this property.
Next we give the bounds of the negativity of the superposition state. The
discussion and conclusion are presented in the end.

The authors in\cite{linden} have shown that two states of high fidelity to
one another may not have the same entanglement, i.e., $|\langle \psi |\phi
\rangle |^{2}\rightarrow 1$ may not generally result in $E(\psi )\rightarrow
E(\phi )$, where $E$ is the von Neumann entropy. For a bipartite pure state $%
|\Phi \rangle _{AB}$ the von Neumann entropy is defined as
\begin{equation}
E(\Phi _{AB})\equiv S(\mathrm{Tr}_{B}|\Phi \rangle _{AB}\langle \Phi |)=S(%
\mathrm{Tr}_{A}|\Phi \rangle _{AB}\langle \Phi |),  \label{33}
\end{equation}%
where $S(\rho )=-\mathrm{Tr}(\rho \log \rho )$, and the concurrence is
defined as
\begin{equation}
\mathcal{C}(\Phi _{AB})\equiv \sqrt{2\left(1-\mathrm{Tr\rho _{A}^{2}}\right)}%
=\sqrt{2\left(1-\sum\nolimits_{i}\mu _{i}^{2}\right)},  \label{55}
\end{equation}%
where $\rho _{A}=\mathrm{Tr}_{B}|\Phi \rangle _{AB}\langle \Phi|$ with the
eigenvalues $\mu_i$. However, if we employ the concurrence to quantify the
entanglement, $|\langle \psi |\phi \rangle |^{2}\rightarrow 1$ must result
in $\mathcal{C}(\psi )\rightarrow \mathcal{C}(\phi )$. Let us see their
example letting
\begin{equation}
|\phi \rangle _{AB}=|00\rangle,  \label{12}
\end{equation}%
and
\begin{equation}
|\psi \rangle _{AB}=\sqrt{1-\epsilon }|\phi \rangle _{AB}+\sqrt{\frac{%
\epsilon }{d}}[|11\rangle +|22\rangle +\dots +|dd\rangle ].  \label{33}
\end{equation}%
It is obviously true that $E(\phi _{AB})=\mathcal{C}(\phi _{AB})=0$, while
according to \cite{linden} the von Neumann entropy of the state $|\psi
\rangle _{AB}$ is
\begin{equation}
E(\psi _{AB})\approx \epsilon \log _{2}d\rightarrow \infty ,  \label{34}
\end{equation}%
specially when $d$ is as large as we expect. It follows from Eq.(\ref{55})
that the concurrence of the state $|\psi \rangle _{AB}$ give us the result
\begin{equation}
\mathcal{C}^{2}(\psi _{AB})=2\left( 2\epsilon -\epsilon ^{2}-\frac{\epsilon
^{2}}{d}\right) \rightarrow 0,  \label{a}
\end{equation}%
when $\epsilon $ is adequately small. By contrast to $E(\psi _{AB})$ in Eq.(%
\ref{34}), $\mathcal{C}^{2}(\psi _{AB})$ in Eq.(\ref{a}) is independent of $%
d $. Note that when $\epsilon $ is small the two states have high fidelity $%
|\langle \psi |\phi \rangle |^{2}=1-\epsilon \rightarrow 1$. Comparing Eq.(%
\ref{34}) to Eq.(\ref{a}), we can draw a conclusion that if the entanglement
is quantified by the concurrence two states of high fidelity to one another
still have nearly the same entanglement.

It is indeed that the difference of the von Neumann entropy between two pure
states of fixed dimension can be bounded using Fannes' inequality\cite%
{fannes}, while the von Neumann entropy is not a continuous function
and no such bound applies in infinite dimensions. However, as we
will show here, a similar bound still works if the entanglement is
quantified by the concurrence and the concurrence is a continuous
function even in infinite dimensions. In order to explain our above
viewpoint we present the following Theorem which is similar to the
original Fannes' inequality except that the entanglement is
quantified by the concurrence.

\emph{Theorem 1.} Suppose $\rho _{AB}$ and $\sigma_{AB}$ are density
matrices of two bipartite pure states in arbitrary dimensions. For the trace
distance $T(\rho_A,\sigma_A)\equiv \mathrm{Tr}|\rho_A-\sigma_B|$ between $%
\rho_{A}=\mathrm{Tr}_B \rho_{AB}$ and $\sigma_{A}=\mathrm{Tr}_B \sigma_{AB}$
we have
\begin{equation}  \label{d}
|\mathcal{C}^{2}(\rho_{AB})-\mathcal{C}^{2}(\sigma_{AB})|\leq
4T(\rho_{A},\sigma_{A}).
\end{equation}

\emph{Proof.} Let $r_{1}\geq r_{2}\geq \dots \geq r_{d}$ be the eigenvalues
of $\rho _{A}$, in decreasing order, and $s_{1}\geq s_{2}\geq \dots \geq
s_{d}$ be the eigenvalues of $\sigma _{A}$, also in decreasing order.
According to\cite{nielsen}, it follows that
\begin{equation}
\sum_{i}|r_{i}-s_{i}|\leq T(\rho _{A},\sigma _{A}).  \label{e}
\end{equation}%
From the observation of the definition of the concurrence in Eq.(\ref{55}),
we can rewrite the left-hand-side of Eq.(\ref{d}) as
\begin{eqnarray}
\left\vert \mathcal{C}^{2}(\rho _{AB})-\mathcal{C}^{2}(\sigma
_{AB})\right\vert  &=&2\left\vert \sum_{i}(r_{i}^{2}-s_{i}^{2})\right\vert
\notag \\
&\leq &2\sum_{i}\left\vert r_{i}^{2}-s_{i}^{2}\right\vert   \notag \\
&=&2\sum_{i}|r_{i}+s_{i}||r_{i}-s_{i}|  \notag \\
&\leq &4\sum_{i}|r_{i}-s_{i}|.  \label{fe}
\end{eqnarray}%
The second formula is obtained from the observation that $\left\vert
a+b+\cdot \cdot \cdot +k\right\vert \leq \left\vert a\right\vert +\left\vert
b\right\vert +\cdot \cdot \cdot +\left\vert k\right\vert $ for any complex quantities $%
a,b,\cdot \cdot \cdot ,k.$ In the derivation of the last formula we
have taken into account the fact that $r_{i}+s_{i}\leq 2$ since each
eigenvalue of $r_{i}$ and $s_{i} $ is not greater than one.
Combining Eqs.(\ref{e}) and (\ref{fe}) can give Eq.(\ref{d}). Thus
the proof is completed.

From the Theorem 1 it can be seen that the difference of the
concurrences of two pure states is a function of fidelity and can be
bounded by Eq.(\ref{d}). What's more, by contrast to the von Neumann
entropy\cite{linden} the concurrence is a continuous function and
such a bound still works in infinite dimensions. Note that whether a
similar bound in Eq.(\ref{d}) holds for the negativity is still
open. In the next paragraphs we are devoted to deducing the bounds
on the negativity of any bipartite pure state as a superposition of
two terms $|\Gamma\rangle_{AB}=\alpha|\Psi\rangle+\beta
|\Phi\rangle$.

Before embarking on this study, we first recall some basic definitions of
the negativity. As for detecting entangled state in higher-dimensional
Hilbert space, Peres-Horodecki criterion based on partial transpose\cite{pe,
ho} is a convenient method. Given a density matrix $\rho$ in a bipartite
pure system of $A$ and $B$, the partial transpose with respect to $A$
subsystem is described by $(\rho^{T_{A}})_{ij,kl}=(\rho)_{kj,il}$ and the
negativity is defined as
\begin{equation}  \label{h}
\mathcal{N}=\frac{1}{2}(\|\rho^{T_{A}}\|-1).
\end{equation}
The trace norm $\|R\|$ is given by $\|R\|=\mathtt{Tr}\sqrt{RR^{\dagger}}$.
Note that $\mathcal{N}>0$ is the necessary and sufficient condition for
entangled bipartite pure states.

There are two key ingredients to obtain the bounds of the negativity for
bipartite superposition pure states. One is that the negativity can be
expressed by means of Schmidt coefficients of a pure state. Suppose that a
pure $m\otimes n(m\leq n)$ quantum state has the standard Schmidt form $%
|\psi\rangle_{AB}=\sum_{i}\sqrt{\mu_{i}}|a_{i}b_{i}\rangle$, where $\sqrt{%
\mu_{i}}(i=1,\cdots, m)$ are the Schmidt coefficients, $a_{i}$ and $b_{i}$
are the orthogonal basis in $\mathcal{H}_{A}$ and $\mathcal{H}_{B}$,
respectively. For the pure bipartite state we can derive $%
\|\rho^{T_{A}}\|=\left(\sum_{i}\sqrt{\mu_{i}}\right)^{2}$\cite{fei}, and
therefore Eq.(\ref{h}) can be reexpressed as
\begin{equation}  \label{j}
\mathcal{N}=\frac{1}{2}\left[\left(\sum_{i}\sqrt{\mu_{i}}\right)^{2}-1\right]%
.
\end{equation}
In order for the later use we can transform Eq.(\ref{j}) into
\begin{equation}  \label{jj}
\left(\sum_{i}\sqrt{\mu_{i}}\right)^{2}=2\mathcal{N}+1.
\end{equation}
The other is the Theorem\cite{horn}, which states that for any two Hermitian
matrix $H$ and $K$ defined in $\mathcal{C}^{n\times n}$,
\begin{equation}  \label{j1}
\mu_{i}(H)+\mu_{1}(K)\leq\mu_{i}(H+K)\leq \mu_{i}(H)+\mu_{n}(K),
\end{equation}
holds, where $\mu_{i}(\cdot)$ are the eigenvalues in increasing order. If $%
\mu_{1}(K)\geq 0$, from Eq.(\ref{j1}) it is easy to check that
\begin{equation}  \label{jh}
\sqrt{\mu_{i}(H)}\leq \sqrt{\mu_{i}(H+K)}\leq \sqrt{\mu_{i}(H)}+\sqrt{%
\mu_{n}(K)},
\end{equation}
holds also. Then Eq.(\ref{jh}) will be used repeatedly in what follows.

For the negativity of the arbitrary superposition state let us first
see the simplest case in which two bipartite states we are
superposing, $\Phi_{1}$ and $\Psi_{1}$, are
biorthogonal\cite{linden}, i.e., $%
\Phi_{1}\Psi_{1}^{\dag}=\Psi_{1}\Phi_{1}^{\dag}=0$\cite{yu}. Since
the matrix representation of a reduced density matrix will be used,
we
explain the corresponding notations in the following. For the pure state $%
|\Phi\rangle_{AB} $ defined in $m\otimes n$ dimensions, generally it can be
considered as a vector: $|\Phi\rangle_{AB}=[a_{00}, a_{01}, \cdots ,a_{0m},
a_{10},a_{11},\cdots, a_{mn}]^{T} $ with the superscript $T$ denoting
transpose operation. With the matrix notation, the reduced density matrix
reads
\begin{equation}  \label{ii}
\rho_{A}=\Phi\Phi^{\dag},
\end{equation}
whose eigenvalues are $\mu_{i}$ appearing in Eq.(\ref{j}).

\emph{Theorem 2.} Suppose that two biorthogonal pure states $\Phi _{1}$ and $%
\Psi _{1}$, which are defined in $m\otimes n(n\leq m)$ dimensions. The
negativity of their superposed states $\Gamma _{1}=\alpha \Phi _{1}+\beta
\Psi _{1}$ with $|\alpha ^{2}|+|\beta |^{2}=1$ satisfies
\begin{widetext}
\begin{equation}\label{k1}
\frac{2|\alpha|^{2}\mathcal{N}(\Phi_{1})+2|\beta|^{2}\mathcal{N}(\Psi_{1})-1}{4}\leq
\mathcal{N}(\alpha\Phi_{1}+\beta \Psi_{1})\leq
\frac{2|\alpha|^{2}\mathcal{\widetilde{N}}(\Phi_{1})+2|\beta|^{2}\mathcal{\widetilde{N}}(\Psi_{1})-1}{4},
\end{equation}
where
\begin{equation}\nonumber
\mathcal{\widetilde{N}}(\Phi_{1})=\mathcal{{N}}(\Phi_{1})+\frac{n|\beta|\sqrt{\mu_{n}(\Psi_1)[2\mathcal{N}(\Phi_1)+1]}}{|\alpha|}+\frac{n^2|\beta|^{2}\mu_{n}(\Psi_1)}{2|\alpha|^{2}},
\end{equation}
and
\begin{equation}\nonumber
\mathcal{\widetilde{N}}(\Psi_{1})=\mathcal{{N}}(\Psi_{1})+\frac{n|\alpha|\sqrt{\mu_{n}(\Phi_1)[2\mathcal{N}(\Psi_1)+1]}}{|\beta|}+\frac{n^2|\alpha|^{2}\mu_{n}(\Phi_1)}{2|\beta|^{2}}.
\end{equation}

\emph{Proof.} From Eq.(\ref{ii}) the reduced density matrix of the
state $\Gamma_{1}$ can read
\begin{equation}\label{j2}
\Gamma_{1}\Gamma_{1}^{\dag}=|\alpha|^{2}\Phi_1\Phi_1^{\dag}+|\beta|^{2}\Psi_1\Psi_1^{\dag}+\alpha\beta^*\Phi_1\Psi_1^{\dag}+\alpha^{*}\beta\Psi_1\Phi_1^{\dag}.
\end{equation}\end{widetext}
The biorthogonal condition with $\Phi_{1}\Psi_{1}^{\dag}=0$ and $%
\Psi_{1}\Phi_{1}^{\dag}=0$ makes Eq.(\ref{j2}) reduce to
\begin{equation}  \label{j3}
\Gamma_{1}\Gamma_{1}^{\dag}=|\alpha|^{2}\Phi_1\Phi_1^{\dag}+|\beta|^{2}%
\Psi_1\Psi_1^{\dag}.
\end{equation}
Substituting Eq.(\ref{j3}) into the left inequality of Eq.(\ref{j1}) we have
\begin{equation}
|\alpha |^{2}\mu _{i}(\Phi _{1}\Phi _{1}^{\dag })+|\beta |^{2}\mu _{1}(\Psi
_{1}\Psi _{1}^{\dag })\leq \mu _{i}(\Gamma _{1}\Gamma _{1}^{\dag }).
\label{j4}
\end{equation}%
Since $\Psi _{1}\Psi _{1}^{\dag }$ is positive semidefinite, $\mu _{1}(\Psi
_{1}\Psi _{1}^{\dag })\geq 0$. Thus Eq.(\ref{j4}) becomes
\begin{equation}
|\alpha |^{2}\mu _{i}(\Phi _{1}\Phi _{1}^{\dag })\leq \mu _{i}(\Gamma
_{1}\Gamma _{1}^{\dag }).  \label{j5}
\end{equation}%
Taking the square root of both sides in Eq.(\ref{j5}) and the sum of $\sqrt{%
\mu _{i}(\cdot )}$ over all index $i$, we have
\begin{equation}
|\alpha |\sum_{i}\sqrt{\mu _{i}(\Phi _{1}\Phi _{1}^{\dag })}\leq \sum_{i}%
\sqrt{\mu _{i}(\Gamma _{1}\Gamma _{1}^{\dag })}.  \label{j6}
\end{equation}%
In a similar way, substituting Eq.(\ref{j3}) into the right inequality of
Eq.(\ref{jh}) and taking the sum of $\sqrt{\mu _{i}(\cdot )}$ over all index
$i$, we have
\begin{equation}  \label{99}
\sum_{i}\sqrt{\mu _{i}(\Gamma _{1}\Gamma _{1}^{\dag }})\leq |\alpha |\sum_{i}%
\sqrt{\mu _{i}(\Phi _{1}\Phi _{1}^{\dag })}+n|\beta |\sqrt{\mu _{n}(\Psi
_{1}\Psi _{1}^{\dag })}.
\end{equation}%
Substituting Eqs.(\ref{j6}) and (\ref{99}) into Eq.(\ref{jj}), respectively,
we can obtain
\begin{eqnarray}
|\alpha |^{2}\mathcal{N}(\Phi _{1})+\frac{|\alpha |^{2}-1}{2} &\leq &%
\mathcal{N}(\alpha \Phi _{1}+\beta \Psi _{1})  \notag \\
&\leq &|\alpha |^{2}{\mathcal{\widetilde{N}}}(\Phi _{1})+\frac{|\alpha
|^{2}-1}{2} .  \label{ee}
\end{eqnarray}
If we replace the matrix $|\alpha|^2\Phi_1\Phi_1^{\dag}$ with $%
|\beta|^2\Psi_1\Psi_1^{\dag}$ in Eqs.(\ref{j5}) and (\ref{j6}), i.e.,
equivalently exchange the matrixes $H$ and $K$ in Eq.(\ref{jh}), finally we
can also obtain
\begin{eqnarray}
|\beta |^{2}\mathcal{N}(\Psi _{1})+\frac{|\beta |^{2}-1}{2} &\leq &\mathcal{N%
}(\alpha \Phi _{1}+\beta \Psi _{1})  \notag \\
&\leq &|\beta |^{2}{\mathcal{\widetilde{N}}}(\Psi _{1})+\frac{|\beta |^{2}-1%
}{2} .  \label{ee1}
\end{eqnarray}
Then combining Eqs.(\ref{ee}) and (\ref{ee1}) gives Eq.(\ref{k1}). Thus the
proof is completed.

Note that the lower bound in Eq.(\ref{k1}) can provide a nonzero value only
when $2|\alpha |^{2}\mathcal{N}(\Phi _{1})+2|\beta |^{2}\mathcal{N}(\Psi
_{1})>1$. Next we show an example to illustrate the validity of our bound.
Consider the state
\begin{equation}
|\phi \rangle _{AB}=\alpha |\varphi \rangle _{AB}+\beta |\psi \rangle _{AB},
\label{j0}
\end{equation}%
with
\begin{equation}
|\varphi \rangle _{AB}=\frac{1}{\sqrt{2}}|00\rangle +\frac{1}{\sqrt{2}}%
|11\rangle ,  \label{j11}
\end{equation}%
\begin{equation}
|\psi \rangle _{AB}=\frac{1}{\sqrt{2}}|22\rangle +\frac{1}{\sqrt{2}}%
|33\rangle ,
\end{equation}%
where $\alpha =\beta =1/\sqrt{2}$. It is easy to check that $|\varphi
\rangle _{AB}$ and $|\psi \rangle _{AB}$ are biorthogonal, $\mathcal{N}%
(|\phi \rangle _{AB})=3/2$, $\mathcal{N}(|\varphi \rangle _{AB})=\mathcal{N}%
(|\psi \rangle _{AB})=1/2$, and $\mu _{4}(|\varphi \rangle _{AB})=\mu
_{4}(|\psi \rangle _{AB})=1/2$. Accordingly from Eq.(\ref{k1}) we obtain the
lower and upper bounds
\begin{equation}
0<\mathcal{N}(|\phi \rangle _{AB})=\frac{3}{2}<4,  \label{j12}
\end{equation}%
which work well.

Finally we directly present the main Theorem of this paper, in which the two
states being superposed can be biorthoganal, orthogonal, or nonorthogonal.

\emph{Theorem 3.} Suppose that two arbitrary normalized pure states $\Phi
_{2}$ with rank $r_{1}$ and $\Psi _{2}$ with rank $r_{2}$, which are defined
in any dimensions. The negativity of their superposed states $\Gamma
_{2}=\alpha \Phi _{2}+\beta \Psi _{2}$ with rank $r_{3}$ and $|\alpha
^{2}|+|\beta |^{2}=1$ satisfies
\begin{widetext}
\begin{equation}
2\Vert \alpha |\Phi _{2}\rangle +\beta |\Psi _{2}\rangle \Vert ^{2}\mathcal{N%
}(\alpha \Phi _{2}+\beta \Psi _{2})\leq 2|\alpha |^{2}\mathcal{\widetilde{N}}%
(\Phi _{2})+2|\beta |^{2}\mathcal{\widetilde{N}}(\Psi _{2})-\Vert \alpha
|\Phi _{2}\rangle +\beta |\Psi _{2}\rangle \Vert ^{2}+1,  \label{l}
\end{equation}%
where
\begin{equation}
\mathcal{\widetilde{N}}(\Phi _{2})=\mathcal{N}(\Phi _{2})+\frac{r|\beta |%
\sqrt{\mu _{n}(\Psi _{2})[2\mathcal{N}(\Phi _{2})+1]}}{|\alpha |}+\frac{%
r^{2}|\beta |^{2}\mu _{n}(\Psi _{2})}{2|\alpha |^{2}},  \notag
\end{equation}%
\begin{equation}
\mathcal{\widetilde{N}}(\Psi _{2})=\mathcal{N}(\Psi _{2})+\frac{r|\alpha |%
\sqrt{\mu _{n}(\Phi _{2})[2\mathcal{N}(\Psi _{2})+1]}}{|\beta |}+\frac{%
r^{2}|\alpha |^{2}\mu _{n}(\Phi _{2})}{2|\beta |^{2}},  \notag
\end{equation}
where $r=\max \{r_{1},r_{2},r_{3}\}$.
\end{widetext}

\emph{Proof.} Consider the matrix
\begin{equation}  \label{kl}
M=|\alpha|^2\Phi_{2}\Phi_{2}^{\dag}+|\beta|^2\Psi_{2}\Psi_{2}^{\dag},
\end{equation}
which can be rewritten as
\begin{equation}  \label{kl1}
M=\frac{\|\Gamma_2\|^2}{2}\widehat{\Gamma}_2 (\widehat{\Gamma}_2)^{\dag}+%
\frac{\|\Gamma_2^{-}\|^2}{2}\widehat{\Gamma}_2^{-} (\widehat{\Gamma}%
_2^{-})^{\dag},
\end{equation}
where $\Gamma_2^{-}=\alpha\Phi_{2}-\beta \Psi_{2}$, $\widehat{\Gamma}%
_2=\Gamma_2/\|\Gamma_2\|$, and $\widehat{\Gamma}_2^{-}=
\Gamma_2^{-}/\|\Gamma_2^{-}\|$. Thus Eqs.(\ref{j1}) shows that

\begin{eqnarray}
&&|\alpha |^{2}\mu _{i}(\Phi _{2}\Phi _{2}^{\dag })+|\beta |^{2}\mu
_{1}(\Psi _{2}\Psi _{2}^{\dag })  \notag \\
&&  \notag \\
&\leq &\mu _{i}(M)\leq |\alpha |^{2}\mu _{i}(\Phi _{2}\Phi _{2}^{\dag
})+|\beta |^{2}\mu _{n}(\Psi _{2}\Psi _{2}^{\dag }),  \label{j4p}
\end{eqnarray}%
and
\begin{eqnarray}
&&\frac{\Vert \Gamma _{2}\Vert ^{2}}{2}\mu _{i}\left( \widehat{\Gamma }_{2}%
\widehat{\Gamma }_{2}^{\dag }\right) +\frac{\Vert \Gamma _{2}^{-}\Vert ^{2}}{%
2}\mu _{1}\left( \widehat{\Gamma }_{2}^{-}(\widehat{\Gamma }_{2}^{-})^{\dag
}\right)  \notag \\
&&  \notag \\
&\leq &\mu _{i}(M)\leq \frac{\Vert \Gamma _{2}\Vert ^{2}}{2}\mu _{i}\left(
\widehat{\Gamma }_{2}\widehat{\Gamma }_{2}^{\dag }\right) +\frac{\Vert
\Gamma _{2}^{-}\Vert ^{2}}{2}\mu _{n}\left( \widehat{\Gamma }_{2}^{-}(%
\widehat{\Gamma }_{2}^{-})^{\dag }\right) .  \label{klj}
\end{eqnarray}%
Since $\mu _{1}(\Psi _{2}\Psi _{2}^{\dag })\geq 0$ and $\mu _{1}\left(
\widehat{\Gamma }_{2}^{-}(\widehat{\Gamma }_{2}^{-})^{\dag }\right) \geq 0$,
observing the left inequality of Eq.(\ref{klj}) and the right inequality in
Eq.(\ref{j4p}) we have
\begin{equation}
\frac{\Vert \Gamma _{2}\Vert }{\sqrt{2}}\sqrt{\mu _{i}\left( \widehat{\Gamma
}_{2}\widehat{\Gamma }_{2}^{\dag }\right) }\leq |\alpha |\sqrt{\mu _{i}(\Phi
_{2}\Phi _{2}^{\dag })}+|\beta |\sqrt{\mu _{n}(\Psi _{2}\Psi _{2}^{\dag })}.
\label{122}
\end{equation}%
Substituting Eqs.(\ref{122}) into Eq.(\ref{jj}) we have
\begin{eqnarray}
&&\Vert \alpha |\Phi _{2}\rangle +\beta |\Psi _{2}\rangle \Vert ^{2}\mathcal{%
N}(\alpha \Phi _{2}+\beta \Psi _{2})  \notag \\
&&  \notag \\
&\leq &2|\alpha |^{2}\mathcal{\widetilde{N}}(\Phi _{2})-\frac{\Vert \alpha
|\Phi _{2}\rangle +\beta |\Psi _{2}\rangle \Vert ^{2}}{2}+|\alpha |^{2}.
\label{er}
\end{eqnarray}%
Likewise, if we replace the two matrixes $|\alpha |^{2}\Phi _{2}\Phi
_{2}^{\dag }$ with $|\beta |^{2}\Psi _{2}\Psi _{2}^{\dag }$ in Eq.(\ref{j4p}%
), we can obtain
\begin{eqnarray}
&&\Vert \alpha |\Phi _{2}\rangle +\beta |\Psi _{2}\rangle \Vert ^{2}\mathcal{%
N}(\alpha \Phi _{2}+\beta \Psi _{2})  \notag \\
&&  \notag \\
&\leq &2|\beta |^{2}\mathcal{\widetilde{N}}(\Psi _{2})-\frac{\Vert \alpha
|\Phi _{2}\rangle +\beta |\Psi _{2}\rangle \Vert ^{2}}{2}+|\beta |^{2}.
\label{er1}
\end{eqnarray}%
Combining Eqs.(\ref{er}) and (\ref{er1}) gives Eq.(\ref{l}). Thus the proof
is completed.

Since there exists a extra term of the maximal eigenvalue in the second
inequality in Eq.(\ref{klj}), generally it is difficult to achieve the
universal formula for the lower bound of the negativity in this case. But it
is our interest in the future work.

In conclusion, we have shown that if the entanglement is quantified
by the concurrence two pure states of high fidelity to one another
still have nearly the same entanglement and obtained an inequality
that can guarantee that the concurrence is a continuous function
even in infinite dimensions. However, whether the similar property
can apply to the negativity is still open. The bounds on the
negativity of superposed states in terms of those of the states
being superposed were obtained. So far some bounds of the
wildly-studied measures of entanglement like the von Neumann entropy\cite%
{linden}, the concurrence\cite{yu} and the negativity in this paper
for the superposition states have been provided.  In view of that
the concurrence can be directly accessible in laboratory
experiment\cite{wa}, these bounds can find useful applications in
estimating the amount of the entanglement of a given pure state.

\bigskip

The author Y.C.O. was supported from China Postdoctoral Science
Foundation and the author H.F. was supported by 'Bairen' program
NSFC grant and '973' program (2006CB921107).

\end{document}